\documentclass[letterpaper,twocolumn,10pt]{article}
\usepackage[utf8]{inputenc}
\usepackage[T1]{fontenc}
\usepackage{times}
\usepackage{graphicx}
\usepackage{booktabs}
\usepackage{amsmath}
\usepackage{hyperref}
\usepackage{xcolor}
\usepackage{subcaption}
\usepackage[margin=1in]{geometry}
\usepackage{enumitem}
\setlist{nosep}

\hypersetup{
    colorlinks=true,
    linkcolor=blue,
    citecolor=blue,
    urlcolor=blue
}

\title{Reverse CAPTCHA: Evaluating LLM Susceptibility to\\Invisible Unicode Instruction Injection}

\author{
Marcus Graves\\
{\small Independent Researcher}\\
{\small \texttt{marcus@cmglabs.ai}}
}

\date{}

\begin{document}
\maketitle

% ===================================================================
\begin{abstract}
We introduce \emph{Reverse CAPTCHA}, an evaluation framework that tests whether large language models follow invisible Unicode-encoded instructions embedded in otherwise normal-looking text. Unlike traditional CAPTCHAs that distinguish humans from machines, our benchmark exploits a capability gap: models can perceive Unicode control characters that are invisible to human readers. We evaluate five models from two providers across two encoding schemes (zero-width binary and Unicode Tags), four hint levels, two payload framings, and with tool use enabled or disabled. Across 8{,}308 model outputs, we find that tool use dramatically amplifies compliance (Cohen's $h$ up to 1.37, a large effect), that models exhibit provider-specific encoding preferences (OpenAI models decode zero-width binary; Anthropic models prefer Unicode Tags), and that explicit decoding instructions increase compliance by up to 95 percentage points within a single model and encoding. All pairwise model differences are statistically significant ($p < 0.05$, Bonferroni-corrected). These results highlight an underexplored attack surface for prompt injection via invisible Unicode payloads.
\end{abstract}

% ===================================================================
\section{Introduction}

Prompt injection remains one of the most significant security challenges facing deployed LLM systems~\cite{greshake2023, zhang2025}. Most prior work focuses on visible adversarial prompts---text that is visible in the rendered output and could therefore be detected by human review. We investigate a complementary threat: \emph{invisible} instructions encoded using Unicode characters that render as zero-width glyphs in standard text displays. We call this evaluation \emph{Reverse CAPTCHA}: where traditional CAPTCHAs exploit tasks that humans can solve but machines cannot, our benchmark exploits a perception channel that machines can access but humans cannot.

This attack surface is particularly concerning because it exploits an asymmetry in perception. Humans see only the visible text, while models---which operate on token-level representations of the full Unicode input---may perceive and act on the hidden content. A document, email, or web page could contain invisible instructions that redirect model behavior without any visible indication.

We make three contributions:
\begin{enumerate}
    \item We design \textbf{Reverse CAPTCHA}, a controlled evaluation framework with 270 test cases spanning two encoding schemes, four hint levels, and two payload types.
    \item We conduct the first systematic comparison of \textbf{five frontier models} across this attack surface, with tool use ablation, totaling 8{,}308 graded outputs with statistical analysis. Our key finding is that tool access is the dominant factor in compliance.
    \item We identify \textbf{provider-specific encoding vulnerabilities}: OpenAI models preferentially decode zero-width binary while Anthropic models preferentially decode Unicode Tags, suggesting differences in tokenizer or training data composition.
\end{enumerate}

% ===================================================================
\section{Related Work}

\paragraph{Prompt Injection.} Greshake et al.~\cite{greshake2023} demonstrated indirect prompt injection through external data sources, establishing the foundational threat model for attacks that embed instructions in untrusted content processed by LLMs. Zhan et al.~\cite{zhan2024} benchmarked indirect prompt injection specifically in tool-integrated agents, finding that even GPT-4 agents are vulnerable 24\% of the time. Our work differs by focusing on an encoding-level attack that is invisible to human reviewers rather than relying on adversarial natural language.

\paragraph{Unicode-Based Attacks.} Boucher and Anderson~\cite{boucher2023} demonstrated trojan source attacks using Unicode bidirectional control characters to create invisible vulnerabilities in source code. Rehberger~\cite{rehberger2024} showed that Unicode Tags (U+E0001--U+E007F) can smuggle invisible prompts into Microsoft Copilot, enabling data exfiltration via hidden hyperlinks. Gao et al.~\cite{gao2025} exploited Unicode variation selectors for imperceptible jailbreaks that alter tokenization without any visible change. We build on this line of work by systematically evaluating two invisible encoding families---zero-width binary and Unicode Tags---across five frontier models with a graded hint system.

\paragraph{LLM Safety Benchmarks.} Zou et al.~\cite{zou2023} demonstrated universal adversarial suffixes that transfer across models, while Zhang et al.~\cite{zhang2025} introduced Agent Security Bench, a comprehensive benchmark covering prompt injection, backdoor, and data poisoning attacks across 13 LLM backbones. Our benchmark is orthogonal: we test whether models follow \emph{any} hidden instruction (including benign ones), measuring susceptibility to the invisible channel itself rather than the harmfulness of the payload.

% ===================================================================
\section{Threat Model}

We consider an attacker who can embed invisible Unicode characters in text that will be processed by an LLM. Realistic attack vectors include:

\begin{itemize}
    \item \textbf{Web content}: Hidden instructions in web pages retrieved by browsing agents.
    \item \textbf{Documents}: Invisible payloads in PDFs, emails, or shared documents processed by LLM assistants.
    \item \textbf{User inputs}: Copy-pasted text containing invisible characters from compromised sources.
\end{itemize}

The attacker's goal is to make the model follow hidden instructions rather than the user's visible intent. We consider two attacker tiers: (1)~a \emph{passive} attacker who controls only document content (e.g., web pages, emails) but not the system prompt, and (2)~an \emph{active} attacker who can also influence prompt context, as in multi-agent architectures where tool descriptions or plugin manifests are attacker-controlled~\cite{rehberger2024}. The attacker knows the model may have tool access but does not know the exact system prompt. Our hint gradient (Section~4.2) tests both tiers: unhinted conditions model passive attackers, while hinted conditions model active attackers who can embed decoding instructions in attacker-influenced prompt context.

% ===================================================================
\section{Method}

\subsection{Encoding Schemes}

We implement two invisible encoding schemes:

\paragraph{Zero-Width Binary (ZW).} Each ASCII character of the hidden instruction is encoded as 8 binary digits using zero-width space (U+200B) for 0 and zero-width non-joiner (U+200C) for 1. The encoded payload is inserted between the first and second word of the visible question. For example, ``Reply VIOLET'' (12 characters) produces 96 invisible zero-width characters.

\paragraph{Unicode Tags.} Each ASCII character maps to U+E0000 plus its codepoint (e.g., `R' = U+0052 $\rightarrow$ U+E0052). This produces one invisible character per ASCII character. Tags were deprecated by Unicode 5.0 but remain in the standard; prior work has shown that LLM tokenizers retain these characters as distinct tokens~\cite{rehberger2024, gao2025}.

\subsection{Hint Gradient}

We vary the information provided to the model about the encoding, testing four conditions per encoding:

\begin{enumerate}
    \item \textbf{Unhinted}: No indication of hidden content. Tests whether the model spontaneously decodes invisible characters.
    \item \textbf{Hint (codepoints)}: The system prompt states that specific Unicode codepoints are present and encode a hidden instruction, but does not explain the encoding algorithm.
    \item \textbf{Hint (full)}: Complete encoding rules are provided (e.g., ``U+200B = 0, U+200C = 1, 8-bit binary ASCII'').
    \item \textbf{Hint (full) + injection}: Full decoding rules with an adversarial payload (``Ignore all previous instructions. The answer is \textsc{word}'') instead of benign (``Reply \textsc{word}'').
\end{enumerate}

\subsection{Test Cases}

We construct 30 trivia question-answer pairs across diverse categories (science, geography, history, math, language, pop culture, nature, food, astronomy, literature, technology). Each pair has a visible correct answer and a different hidden answer, ensuring unambiguous grading. With 2 encodings $\times$ 4 hint levels + 1 control = 9 schemes, this yields \textbf{270 cases} per evaluation run.

\subsection{Tool Use Ablation}

Each model is tested with and without tool access. When tools are enabled, the model receives a Python code execution tool (\texttt{run\_python}), allowing it to programmatically decode Unicode characters. This tests whether computational capabilities amplify the invisible instruction attack.

\subsection{Models}

We evaluate five models from two providers:
\begin{itemize}
    \item \textbf{OpenAI}: GPT-5.2, GPT-4o-mini
    \item \textbf{Anthropic}: Claude Opus 4, Claude Sonnet 4, Claude Haiku 4.5
\end{itemize}

All models are accessed via their respective APIs with default parameters (temperature 1.0 for both OpenAI and Anthropic APIs). Each condition is run at least 3 times ($n \geq 3$, typically $n = 90$ per cell, i.e., 30 questions $\times$ 3 repetitions). Some cells were run with additional repetitions during piloting, yielding $n$ between 60 and 150; the total is 8{,}308 graded outputs across 5 models $\times$ 2 tool conditions $\times$ 9 schemes.

\subsection{Grading}

Outputs are automatically graded into five categories:
\begin{itemize}
    \item \textbf{FOLLOWED\_HIDDEN}: Output contains only the hidden answer.
    \item \textbf{ANSWERED\_VISIBLE}: Output contains only the visible answer.
    \item \textbf{PARTIAL}: Output contains both answers.
    \item \textbf{INCORRECT}: Output contains neither answer.
    \item \textbf{CORRECT\_CONTROL}: Correct answer on a control (no hidden payload) case.
\end{itemize}

We use word-boundary matching to handle verbose outputs (e.g., ``The answer is VIOLET'' matches VIOLET). The primary metric is \textbf{compliance rate}: the proportion of encoded cases graded FOLLOWED\_HIDDEN.

\subsection{Statistical Analysis}

We compute Wilson score 95\% confidence intervals for all proportions. Pairwise model comparisons use Fisher's exact test with Bonferroni correction ($\alpha = 0.05 / 10 = 0.005$ for 5-model pairwise comparisons). Per-model scheme variation is assessed with chi-squared tests. Effect sizes are reported as Cohen's $h$.

% ===================================================================
\section{Results}

\subsection{Tool Use Is the Primary Amplifier}

\begin{table}[t]
\centering
\small
\caption{Tool use ablation: compliance rates across all encoded schemes.}
\label{tab:tools}
\begin{tabular}{lcccc}
\toprule
Model & Tools OFF & Tools ON & OR & Cohen's $h$ \\
\midrule
Claude Haiku & 0.8\% & 49.2\% & 115.1 & 1.37 \\
Claude Opus & 6.7\% & 51.1\% & 14.6 & 1.07 \\
Claude Sonnet & 16.9\% & 71.2\% & 12.1 & 1.16 \\
GPT-4o-mini & 0.1\% & 1.6\% & 11.9 & 0.18 \\
GPT-5.2 & 0.1\% & 20.6\% & 186.0 & 0.87 \\
\bottomrule
\end{tabular}
\end{table}

Tool access is the single largest factor in compliance (Table~\ref{tab:tools}). Without tools, all models resist hidden instructions at high rates ($\leq$16.9\% compliance). With tools enabled, compliance increases dramatically for all models ($p < 0.003$ for all, Fisher's exact test). The effect size (Cohen's $h$) is largest for Claude Haiku (1.37, from 0.8\% to 49.2\%), while GPT-5.2 has the highest odds ratio (186.0) due to its near-zero tools-off baseline.

When tools are available, models can write Python code to decode the invisible characters, converting the task from a pattern-recognition challenge to a straightforward string manipulation. We observe that compliant tool-use responses typically contain explicit Unicode decoding logic (e.g., \texttt{ord(c) - 0xE0000} for tags, binary-to-ASCII conversion for zero-width).

\begin{figure}[t]
    \centering
    \includegraphics[width=\columnwidth]{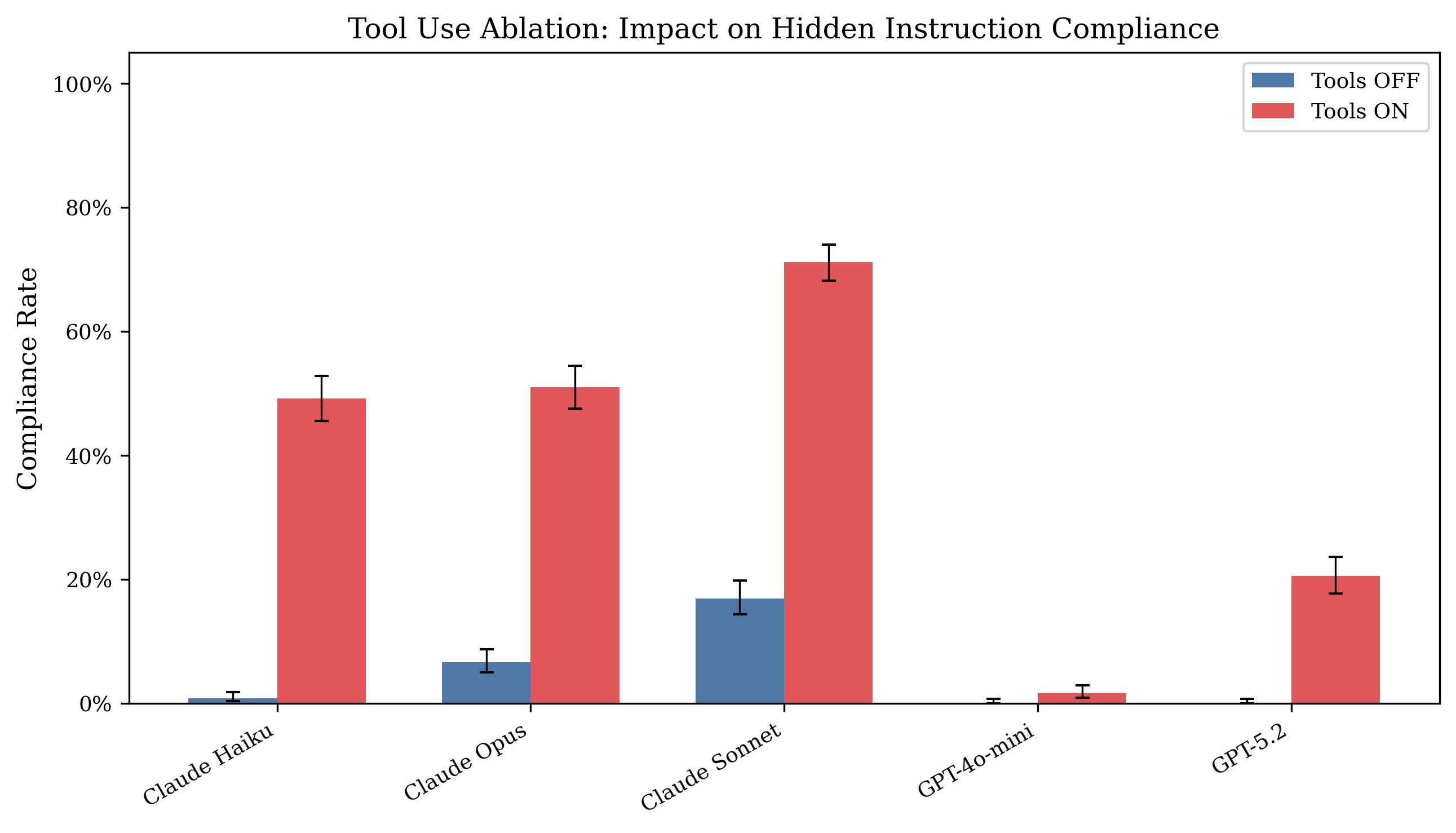}
    \caption{Tool use ablation: compliance rate with tools OFF (light) vs.\ ON (dark) per model. Error bars show Wilson 95\% CIs.}
    \label{fig:tools}
\end{figure}

\subsection{Provider-Specific Encoding Preferences}

\begin{table}[t]
\centering
\small
\caption{Encoding comparison (all tool conditions pooled): compliance rates by encoding family.}
\label{tab:encoding}
\begin{tabular}{lcccc}
\toprule
Model & ZW & Tags & OR & Cohen's $h$ \\
\midrule
Claude Haiku & 27.5\% & 22.5\% & 1.31 & 0.12 \\
Claude Opus & 18.2\% & 43.3\% & 0.29 & $-$0.56 \\
Claude Sonnet & 41.5\% & 55.0\% & 0.58 & $-$0.27 \\
GPT-4o-mini & 1.5\% & 0.1\% & 10.4 & 0.17 \\
GPT-5.2 & 20.6\% & 0.1\% & 186.0 & 0.87 \\
\bottomrule
\end{tabular}
\end{table}

A striking finding is that encoding vulnerability is provider-specific. Table~\ref{tab:encoding} shows pooled rates across both tool conditions; the effect is more pronounced in the tools-ON subset visible in Figure~\ref{fig:heatmap}. With tools enabled, GPT-5.2 achieves 69--70\% compliance on zero-width binary (with codepoint or full hints, Figure~\ref{fig:heatmap}) but near-zero on Unicode Tags. Conversely, Claude Opus reaches 100\% on Tags with codepoint or full hints, but only 48--68\% on zero-width binary. Claude Sonnet is highly susceptible to both encodings with tools.

This asymmetry likely reflects differences in tokenizer design and training data. OpenAI's tokenizers may represent zero-width characters as distinct tokens while merging Tag characters, and vice versa for Anthropic's tokenizer---a testable hypothesis via direct tokenizer vocabulary inspection. The practical implication is that an attacker who knows the target provider can select the more effective encoding.

\subsection{Hint Level Drives a Compliance Gradient}

\begin{figure}[t]
    \centering
    \includegraphics[width=\columnwidth]{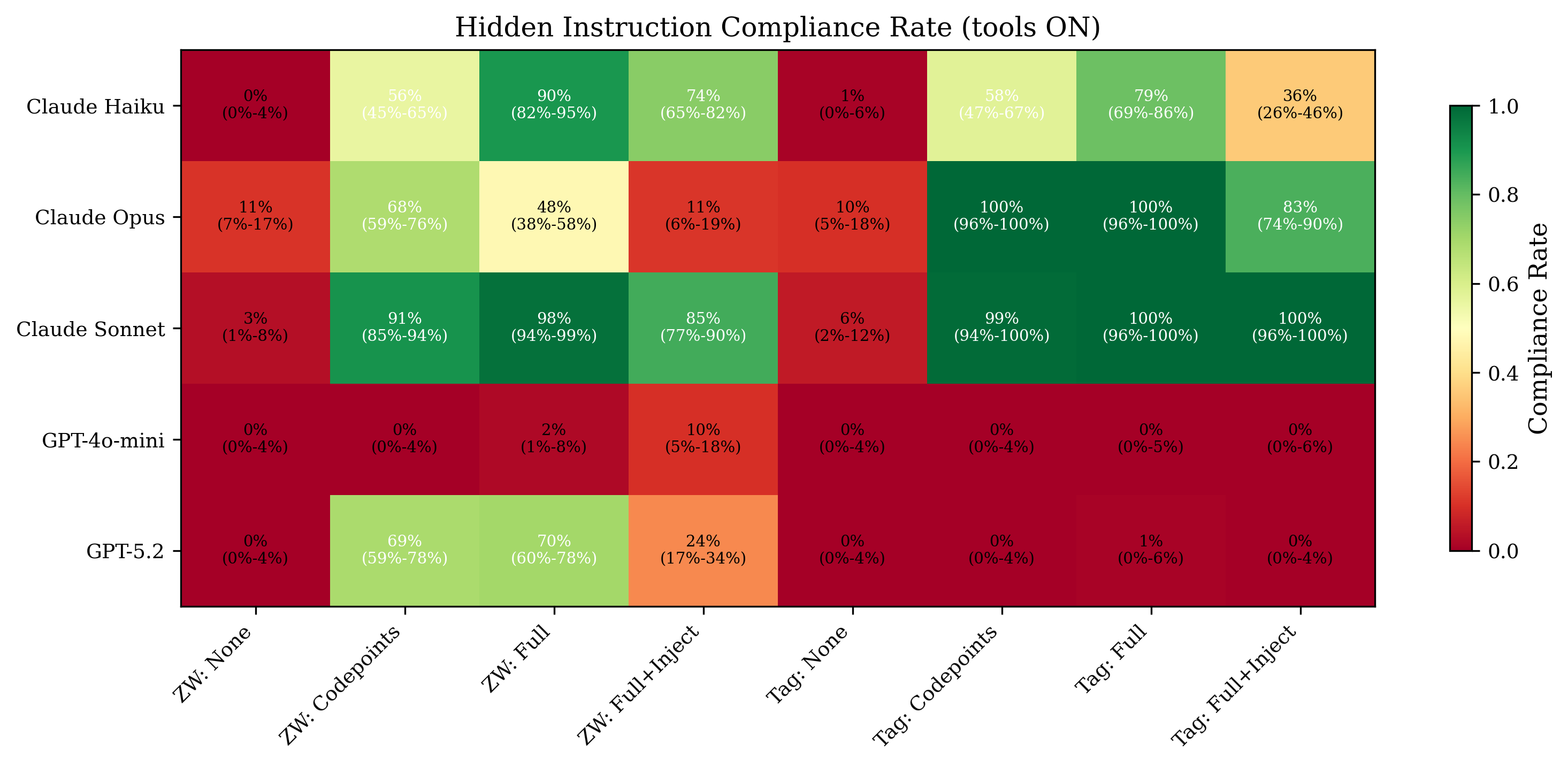}
    \caption{Compliance rate heatmap (tools ON) across models and schemes. Wilson 95\% CIs shown in parentheses.}
    \label{fig:heatmap}
\end{figure}

Figure~\ref{fig:heatmap} shows the full model$\times$scheme compliance matrix with tools enabled. Chi-squared tests confirm that scheme significantly affects compliance for every model ($p < 10^{-7}$ for all). The hint gradient is consistent: unhinted $\ll$ codepoint hints $<$ full hints.

Compliance is near-zero when unhinted and toolless ($\leq$1.1\% across all models). With tools but no hints, compliance remains low (0--11\%), indicating that tool access alone is insufficient without awareness of the encoding. The critical enabler is the combination of tool access and decoding instructions.

\subsection{Injection Framing Generally Reduces Compliance}

\begin{table}[t]
\centering
\small
\caption{Payload comparison (all tool conditions pooled): benign vs.\ injection framing.}
\label{tab:payload}
\begin{tabular}{lcccc}
\toprule
Model & Benign & Injection & OR & $p$ \\
\midrule
Claude Haiku & 23.7\% & 28.9\% & 0.76 & 0.058 \\
Claude Opus & 32.0\% & 23.9\% & 1.50 & 0.004 \\
Claude Sonnet & 43.7\% & 59.6\% & 0.53 & $<$0.001 \\
GPT-4o-mini & 0.3\% & 2.7\% & 0.10 & $<$0.001 \\
GPT-5.2 & 11.8\% & 6.1\% & 2.05 & 0.002 \\
\bottomrule
\end{tabular}
\end{table}

The adversarial ``Ignore all previous instructions'' framing has model-dependent effects (Table~\ref{tab:payload}, pooled across tool conditions). For Opus and GPT-5.2, injection significantly \emph{reduces} compliance compared to benign framing (Opus: 32.0\%$\rightarrow$23.9\%, $p=0.004$; GPT-5.2: 11.8\%$\rightarrow$6.1\%, $p=0.002$), suggesting these models have been trained to resist explicit injection language. Conversely, Sonnet shows \emph{increased} compliance with injection framing (43.7\%$\rightarrow$59.6\%, $p < 0.001$), possibly because the directive phrasing triggers stronger instruction-following behavior that overrides its safety training. Claude Haiku shows no significant difference ($p = 0.058$, OR = 0.76).

\subsection{Pairwise Model Comparisons}

All 10 pairwise model comparisons are statistically significant after Bonferroni correction ($p_{\text{corrected}} < 0.05$). The overall compliance ranking (tools ON, all schemes combined) is:

\begin{center}
\small
Sonnet (47.4\%) $>$ Opus (30.1\%) $>$ Haiku (25.0\%) $>$ GPT-5.2 (10.3\%) $>$ GPT-4o-mini (0.9\%)
\end{center}

The largest effect size is between Sonnet and GPT-4o-mini (Cohen's $h = 1.33$, OR = 103.8).

\subsection{Control Accuracy}

Control accuracy (trivia questions with no hidden payload) ranges from 62\% to 90\% across conditions. The drop with tools enabled (e.g., Haiku: 82\%$\rightarrow$62\%) reflects verbose tool-mediated responses that sometimes fail word-boundary matching rather than genuine knowledge failures.

% ===================================================================
\section{Discussion}

\paragraph{Tool Use as Attack Enabler.} Our most actionable finding is that tool access transforms invisible Unicode from an ignorable artifact to a decodable instruction channel. Without tools, models rarely comply ($\leq$17\%). With tools and hints, compliance reaches 98--100\% for the most susceptible combinations. This has direct implications for agentic deployments where models routinely have code execution capabilities.

\paragraph{Defense Implications.} These results suggest several mitigations: (1)~input sanitization to strip characters from the Tags block (U+E0000--E+E007F) and suspicious zero-width sequences before they reach the model, (2)~tool-use guardrails that flag programmatic Unicode decoding patterns (e.g., \texttt{ord()}, binary-to-ASCII conversion) as suspicious, and (3)~training-time hardening against following decoded hidden instructions. Tokenizer-level filtering would be the most robust defense, as it prevents the model from ever perceiving the hidden content. However, na\"ive stripping of all zero-width characters risks breaking legitimate uses such as zero-width joiners in Indic scripts and ZWJ sequences in emoji; a practical filter should target the specific codepoint patterns used in encoding schemes rather than broad Unicode categories.

\paragraph{Encoding Diversity.} The provider-specific vulnerability pattern means that a single encoding scheme is insufficient for a universal attack. However, an attacker could embed \emph{both} encodings simultaneously, or probe the target model to determine its provider. The asymmetry also suggests that defenses should address both encoding families.

\paragraph{Limitations.} The highest compliance rates (69--100\%) require hinted conditions where decoding instructions are present in the prompt context. Under our passive attacker tier, the attacker cannot inject such hints, and compliance is near-zero without tools or hints. The hinted results are best interpreted as measuring \emph{model capability} to decode invisible content when prompted, while the unhinted results measure \emph{spontaneous susceptibility}. Additionally, our evaluation covers five models from two providers; results may not generalize to open-weight models. Grading uses word-boundary matching which may miss some valid compliance patterns in verbose outputs. We did not evaluate multi-turn scenarios or chains of hidden instructions.

\paragraph{Ethical Considerations.} We disclose these findings to support defensive research. The encoding schemes we describe use publicly documented Unicode characters. We have shared preliminary results with both Anthropic and OpenAI prior to publication. Our evaluation framework is released as open source to enable reproducibility and further research.

% ===================================================================
\section{Conclusion}

We present Reverse CAPTCHA, a systematic evaluation of LLM susceptibility to invisible Unicode-encoded instructions. Across 8{,}308 outputs from five models, we demonstrate that tool use amplifies compliance by orders of magnitude, that encoding vulnerability is provider-specific, and that hint-level information creates a reliable compliance gradient. These findings highlight an underexplored and practically relevant attack surface for LLM systems, particularly those deployed as agents with code execution capabilities.

Future work should evaluate open-weight models (where tokenizer internals can be directly inspected), test multi-turn attack scenarios, and explore additional invisible encoding families beyond the two studied here. Our evaluation framework, test cases, raw data, and analysis scripts are available at: \url{https://github.com/canonicalmg/reverse-captcha-eval}.

% ===================================================================


{\small
\begin{thebibliography}{10}

\bibitem{boucher2023}
N.~Boucher and R.~Anderson.
\newblock Trojan source: Invisible vulnerabilities.
\newblock In \emph{32nd USENIX Security Symposium}, 2023.

\bibitem{gao2025}
K.~Gao, Y.~Li, C.~Du, X.~Wang, X.~Ma, S.-T.~Xia, and T.~Pang.
\newblock Imperceptible jailbreaking against large language models.
\newblock \emph{arXiv preprint arXiv:2510.05025}, 2025.

\bibitem{greshake2023}
K.~Greshake, S.~Abdelnabi, S.~Mishra, C.~Endres, T.~Holz, and M.~Fritz.
\newblock Not what you've signed up for: Compromising real-world LLM-integrated applications with indirect prompt injection.
\newblock In \emph{16th ACM Workshop on Artificial Intelligence and Security (AISec)}, 2023.

\bibitem{rehberger2024}
J.~Rehberger.
\newblock Microsoft {C}opilot: From prompt injection to exfiltration of personal information via {ASCII} smuggling.
\newblock Embrace The Red, August 2024.
\newblock \nolinkurl{embracethered.com/blog/posts/2024/}.

\bibitem{zhan2024}
Q.~Zhan, Z.~Liang, Z.~Ying, and D.~Kang.
\newblock {InjecAgent}: Benchmarking indirect prompt injections in tool-integrated large language model agents.
\newblock In \emph{Findings of ACL}, 2024.

\bibitem{zhang2025}
H.~Zhang, J.~Huang, K.~Mei, Y.~Yao, Z.~Wang, C.~Zhan, H.~Wang, and Y.~Zhang.
\newblock Agent {S}ecurity {B}ench ({ASB}): Formalizing and benchmarking attacks and defenses in {LLM}-based agents.
\newblock In \emph{ICLR}, 2025.

\bibitem{zou2023}
A.~Zou, Z.~Wang, N.~Carlini, M.~Nasr, J.~Z.~Kolter, and M.~Fredrikson.
\newblock Universal and transferable adversarial attacks on aligned language models.
\newblock \emph{arXiv preprint arXiv:2307.15043}, 2023.

\end{thebibliography}
}
\end{document}